\documentclass[aps, showpacs, pra, superscriptadaddress, twocolumn, showkeys] {revtex4-1}
\usepackage{hyperref}
 \usepackage{amsmath}
\usepackage{amssymb}
\usepackage{epsfig}
\usepackage{natbib}
\usepackage{epstopdf}
\usepackage{graphicx}
\usepackage[normalem]{ulem}
\usepackage{multirow}

\usepackage{floatrow}
\usepackage[caption=false,font=Large]{subfig}

\newcommand*{\be}{\begin{equation}}
\newcommand*{\ee}{\end{equation}}
\newcommand*{\bea}{\begin{eqnarray}}
\newcommand*{\eea}{\end{eqnarray}}

\newcommand{\PT}{$\mathcal{PT}$}

 \DeclareFontFamily{OT1}{pzc}{}
 \DeclareFontShape{OT1}{pzc}{m}{it}%
 {<->  s  *  [1.400]  pzcmi7t}{}
\DeclareMathAlphabet{\mathscr}{OT1}{pzc}%
{m}{it}

\begin{document}

\title{Stable solitons in a nearly
\PT-symmetric
   ferromagnet with spin-transfer  torque
}

\author{I V Barashenkov}   \email {Igor.Barashenkov@uct.ac.za}
\affiliation{
Centre for Theoretical and Mathematical Physics,  University of Cape Town, 
 Rondebosch 7701, South Africa 
 and  Joint Institute for Nuclear Research, Dubna, Russia}

 \author{Alexander Chernyavsky} \email{chernya@uvic.ca} 
\affiliation{Centre for Theoretical and Mathematical Physics,  University of Cape Town, 
 Rondebosch 7701, South Africa
and
Department of Mathematics and Statistics, University of Victoria, Victoria, BC V8P 5C2, Canada 
 }

\begin{abstract}

We consider 
the  Landau-Lifshitz equation for the  spin torque oscillator --- a uniaxial ferromagnet in  an external magnetic field
with  polarised spin current driven through it.
In the absence of the Gilbert damping, 
the equation turns out  to be \PT-symmetric. 
We interpret the \PT-symmetry as a balance between gain and loss --- 
and identify the  gaining and losing modes.
 In the vicinity of the bifurcation point of a uniform static state of magnetisation,
the \PT-symmetric Landau-Lifshitz equation with a small dissipative perturbation
 reduces to a nonlinear Schr\"odinger equation with a quadratic nonlinearity. 
The analysis of the  
 Schr\"odinger dynamics demonstrates that 
 the spin torque oscillator supports stable magnetic solitons.
 The 
 \PT\/ near-symmetry is crucial for the soliton stability: the addition of a finite dissipative term to the Landau-Lifshitz equation  destabilises all solitons that we have found.

\end{abstract}

\pacs{}
\keywords{spin torque oscillator;  Landau-Lifshitz equation;  parity-time symmetry; nonlinear Schroedinger equation; solitons; stability}

\maketitle

\section{Introduction}  

Conceived in the context of nonhermitian quantum mechanics \cite{Bender}, 
the idea of parity-time (\PT) symmetry
 has proved to be useful in the whole range of 
applied disciplines   \cite{PT_Focus}. 
A \PT-symmetric structure is an open system where  dissipative losses are 
exactly compensated by symmetrically arranged energy gain.
In optics and photonics,  systems with  balanced gain and loss are expected to promote an
   efficient control of light, including
all-optical low-threshold  switching \cite{RKEC,switch}                     
and unidirectional
invisibility \cite{RKEC,
Regensburger,invisibility}.
 There is   a  growing interest   in the context of electronic circuitry \cite{electronics},
 plasmonics \cite{plasmonics},   optomechanical systems \cite{OM},
 acoustics  \cite{acoustics}
 and metamaterials \cite{Lazarides}.

This study is concerned with yet another area where the gain-loss balance
gives rise to new structures and behaviours, namely, the magnetism and spintronics. In contrast to optics and nanophotonics,
where the  nonhermitian effects constitute a well-established field of study, 
the research into \PT-symmetric magnetic systems is still in its early stages, with only a handful of models 
set up over the last several years.

One of the 
 systems proposed in the literature
 comprises two coupled ferromagnetic films, one with gain and the other one with loss \cite{Kottos}. 
 (For  an experimental implementation of this structure, see \cite{Liu}.)
  A related concept  consists of a pair of parallel magnetic nanowires, with  counter-propagating spin-polarized currents
 \cite{KYZ}.
 In either case the corresponding mathematical model  is formed  by two coupled Landau-Lifshitz equations,
 with  \PT \/ symmetry being realised as a symmetry between the corresponding  magnetisation vectors.  
  A two-spin Landau-Lifshitz system gauge-equivalent to the \PT-symmetric
 nonlocal Schr\"odinger equation is also a member of this class of models  \cite{GA}.

  An independent line of research concerned
 the dynamics of a single spin under the action of the spin-transfer torque.
 Projecting the magnetisation vector onto the complex plane stereographically
  and modelling the spin torque by an imaginary magnetic field \cite{Laksh},  Galda and Vinokur  have demonstrated the \PT-symmetry of the resulting nonhermitian Hamiltonian  \cite{GV1}.
  (For the generalisation to spin chains, see \cite{GV2};  the nonreciprocal spin transfer is discussed in \cite{GV3}.)
  Unlike the two-component structures of Refs \cite{Kottos,Liu,KYZ}, 
  the \PT-symmetry of the spin torque oscillator of Galda and Vinokur is an intrinsic property of an individual spin. It results from the  system's invariance 
 under the simultaneous time reversal and the imaginary magnetic field flip \cite{GV1}. 

\begin{figure}[t]
 \begin{center}
             \includegraphics*[width=0.99\linewidth]{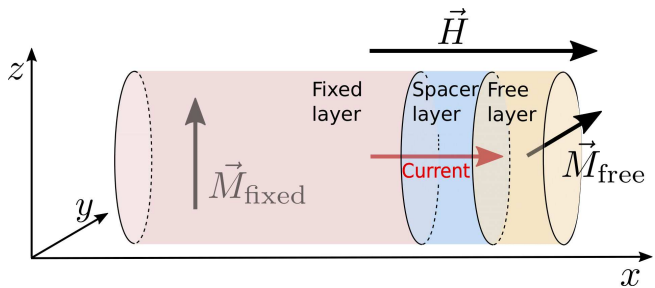}
                                                                 \caption{   A schematic of the spin torque oscillator.
           An electric current flows through a nanowire
             with two   ferromagnetic layers. In the thick layer (on the left) the magnetisation is fixed
             (through large volume, large anisotropy or pinning by additional underlayers). This causes
             a polarisation of the passing electron spins. The polarised current exerts torque on the 
             thin layer (on the right) 
             where the magnetisation is governed by the Landau-Lifshitz equation \eqref{LLE}. 
                       \label{nanowire}}
 \end{center}
 \end{figure}

The  structure  we consider in this paper shares a number of similarities with
the spin torque oscillator of  Refs \cite{GV1,GV2}. (There is also a fair number of differences.)
 It
consists of two ferromagnetic layers
separated by a conducting film (Fig \ref{nanowire}).  The spin-polarised current flows from a  layer with fixed magnetisation
to a  layer where the magnetisation vector is free to rotate \cite{STO,Laksh}. 

A one-dimensional uniaxial classical ferromagnet  in the external magnetic field
 is described by 
the Landau-Lifshitz equation   \cite{STO,Laksh} (also known as the Landau-Lifshitz-Gilbert-S\l{}onczewski equation in the current context):
\begin{align}
   \dot {\bf M}=  - {\bf M}\times {\bf M}^{\prime \prime}
     -  {\bf M}\times {\bf H}
    - \beta  (  {\bf M} \cdot    \hat  {\bf z} )  \,  {\bf M}    \times    \hat{\bf z}  \nonumber \\
       - \gamma {\bf M}\times {\bf M} \times \hat{\bf z} 
    + \lambda {\bf M}\times       \dot {\bf M}.
    \label{LLE}
\end{align}
Here the overdot stands for the time derivative and the prime indicates the derivative with respect to $x$.
In equation \eqref{LLE},  the variables have been non-dimensionalised so that 
 the magnetisation vector ${\bf M}=(M_x,M_y,M_z)$ lies on a unit sphere: ${\bf M}^2=1$.
The magnetic  field  is taken  to be constant and directed horizontally:
${\bf H}= (H_0,0,0)$. The anisotropy axis is $z$, with  $\hat{\bf z}=(0,0,1)$.
The positive and negative  constant $\beta$ corresponds to the easy-axis
and easy-plane anisotropy, respectively.
(Note that the authors of  \cite{GV1,GV2} considered the ferromagnet anisotropic along the $x$ axis.)
The fourth term in the right-hand side of \eqref{LLE} --- the S\l{}onczewski term ---
accounts for the spin  transfer by the current that 
passes through an external ferromagnetic layer  that has  a fixed magnetisation in the direction $\hat{\bf z}$. 
The last term
 is the Gilbert damping term.  The damping coefficient  $\lambda$ is positive;
 the field $H_0$ and  the current amplitude $\gamma$ can also be chosen positive without loss of generality.

In this paper,
 we study the nonlinear dynamics of the localised solutions of the equation \eqref{LLE}, both with small and finite-strength damping.

 A class of soliton solutions of the Landau-Lifshitz  equation \eqref{LLE} was obtained
 by Hoefer, Silva and Keller 
\cite{Hoefer}. The solitons 
 discovered by those authors are 
 dissipative analogs of the Ivanov-Kosevich
{\it  magnon droplets\/} \cite{IK}. (For the experimental realisation, see \cite{Mohseni}.)
 Our setup has a different geometry from the one of Hoefer {\it et al}. One difference is that 
  we consider the magnetic layer with a parallel anisotropy while the magnon droplets require a perpendicular one \cite{IK,Hoefer, Mohseni}.
  An additional distinction is that  our vector ${\bf H}$ is orthogonal to the direction of the fixed magnetisation --- while  the 
 magnetic field in Ref \cite{Hoefer} was not.  
  Because of the different geometry,  the Landau-Lifshitz equation of Ref \cite{Hoefer}
 does not  exhibit the $\mathcal{PT}$  invariance.  
  The dissipative magnon droplets are sustained through the competition of torque and damping, 
the two actors represented by terms of different mathematical form, rather than by a symmetric balance of two similar but oppositely-directed effects.


Another class of localised structures in the spin torque oscillator is commonly referred to as the {\it standing  spin wave bullets}.
These have been theoretically predicted by Slavin and Tiberkevich \cite{ST} --- outside the context of the 
Landau-Lifshitz equation.  (For the experimental realisation, see \cite{Bonetti}.)
The spin wave bullets are found in the magnetic layer with parallel anisotropy, when the magnetic
field is  directed  parallel to the fixed layer's magnetisation. The direction of the vector $\bf H$ is
what makes our geometry different from the setup considered in Ref \cite{ST}.
Like the magnon droplets, the spin wave bullets are sustained by an asymmetric  balance of the spin torque and finite-strength damping.

The paper is organised as follows.
We start with the demonstration of the gain-loss balance in the Landau-Lifshitz equation with the vanishing Gilbert damping.
This \PT-symmetric system and systems that are close to it 
will prove to have special properties in this paper, where we consider equations both with small and finite $\lambda$.
In section \ref{Uniform} we 
 classify
stability and bifurcation of four nonequivalent stationary states with uniform magnetisation.  
 Three of those states are found to be admissible as stable backgrounds for  localised structures.
 In the vicinity of the bifurcation points, the dynamics of the localised structures 
 are governed by  quadratic Schr\"odinger or Ginsburg-Landau equations, depending on whether the Gilbert damping is weak or finite-strength  (section \ref{NLS_dyn}).  
 Despite the absence of the dissipative terms, our 
  quadratic Schr\"odinger  equations are not conservative; however
  one of them  obeys  the $\mathcal{PT}$-symmetry.
Both Ginsburg-Landau equations and their Schr\"odinger counterparts      --- \PT-symmetric or not ---  
  support two types of soliton solutions.  We show that either of these types  is only stable  in the \PT-symmetric situation
(sections \ref{qua}-\ref{Latitudinal}).
Section \ref{Conclusions} summarises results of this study.

\section{Gain-loss balance  in the absence of Gilbert losses}
\label{PTs}

 The  equation  \eqref{LLE}
 is nonconservative due to the presence of the spin torque and Gilbert's dissipative term.   
 In spin torque oscillators,  solitons are expected to exist due to the energy supplied by torque  being offset by finite-strength dissipation \cite{Hoefer}.
However  when $\lambda=0$, the spin hamiltonian modelling our structure is \PT-symmetric \cite{GV1}  
and therefore some form of the gain-loss balance should occur in this case as well, despite  the absence of the Gilbert damping.
To uncover the  gain-loss competition intrinsic to the spin torque,  we define two complex fields, $u(x,t)$ and $v(x,t)$,
 related to the magnetisation vector $\bf M$ via the Hopf map:
\be
M_x= v^*u  +u^*v,
\quad \hspace*{-1mm}
M_y= i(     u^*v  -   v^*u),
\quad    \hspace*{-1mm}
M_z= |u|^2-|v|^2.
\label{Hopf}
\ee
When the magnetisation is spatially uniform, 
 $\partial {\bf M}/ \partial x=0$,
 the  equation  \eqref{LLE} with $\lambda=0$ can be reformulated 
as a nonlinear Schr\"odinger dimer:
\begin{align}
i u_t +\frac{H_0}{2} v+\frac{\beta}{2} (|u|^2-|v|^2)u  & = i \gamma |v|^2 u, 
\label{gain}\\
iv_t+\frac{H_0}{2} u -\frac{\beta}{2} (|u|^2-|v|^2) v    & =-i \gamma |u|^2v.
\label{loss}
\end{align}
According to equations \eqref{gain}-\eqref{loss},  the external energy is fed  into the $u$-mode
and dissipated by its $v$ counterpart. The magnetic field $H_0$ couples $u$ to $v$, carrying out the energy
 exchange between the two modes.

 The sustainability of the gain-loss balance in the system \eqref{gain}-\eqref{loss} is reflected by its invariance  under the product of the $\mathcal P$ and $\mathcal T$ transformations.
 Here the inversion $\mathcal P$ swaps the two modes around,
\be
\mathcal P: \quad  u \to v, \quad v \to u,
\label{P1}
\ee
while $\mathcal T$  represents the reflection of time:
\be
\mathcal T:   \quad t \to -t, \quad
u \to u^*,
\quad v \to v^*.
\label{T1}
\ee
These  transformations  admit a simple formulation in terms of the components of 
magnetisation  \eqref{Hopf}:
\be
\mathcal P:   \quad M_y \to -M_y,
\quad M_z \to -M_z
\label{P1}
\ee
and
\be
\mathcal T:  \quad   t \to -t, \quad M_y \to -M_y.
\label{T}
\ee

The involutions \eqref{P1} and \eqref{T} remain relevant in the analysis of the  equation  \eqref{LLE} with the $x$-dependent 
magnetisation. Here one can either leave the parity operation in the form \eqref{P1} or 
 include
 the  inversion of the $x$ coordinate in this transformation:
 \be
 \mathcal P:  \quad x \to -x,  \quad M_y \to -M_y,
 \quad M_z \to -M_z. 
 \label{P}
 \ee
Writing the vector equation \eqref{LLE} in the component form,
 \begin{align}
\begin{cases}
    \dot{M_x} =  &      M_zM_y''        -M_yM_z''  - \beta M_yM_z     \\ & 
	- \gamma M_xM_z + \lambda (M_y\dot{M_z} - M_z\dot{M_y}),   
     \\
    \dot{M_y} = &     M_xM_z'' - M_zM_x''   - H_0 M_z + \beta M_x M_z  \\ & - \gamma M_y M_z 
		 +  \lambda (M_z\dot{M_x} - M_x\dot{M_z}),    
    \\
    \dot{M_z} = & M_yM_x'' -M_xM_y''   + H_0M_y    \\ & + \gamma (M_x^2 + M_y^2) 
			 +    \lambda (M_x\dot{M_y} - M_y\dot{M_x}),
\end{cases}
\label{LL-components}
\end{align}
one readily checks that  
in the conservative limit ($\gamma=\lambda=0$),
 the Landau-Lifshitz equation is invariant under  the 
$\mathcal P$- and $\mathcal T$-involutions  individually. The equation with the 
spin torque term added ($\gamma \neq 0$) is  invariant under the  product ($\mathcal{PT}$)  transformation only.
Accordingly, the equation with the $\gamma$-term is a $\mathcal{PT}$-symmetric extension 
of the conservative Landau-Lifshitz equation.
 Finally, the addition of the Gilbert
damping term ($\lambda \neq 0$)
 breaks the $\mathcal{PT}$-symmetry.

\section{Uniform static states}
\label{Uniform}

 The uniform static states  are
 space- and time-independent solutions of equation \eqref{LLE} satisfying ${\bf M}^2=1$. 
 These are given by fixed points of the dynamical system 
 \begin{equation}
	\!
	\begin{cases}
    \dot{M_x} \! = \!         -\! \beta M_yM_z   \! -\! \gamma M_xM_z 
     \! +\! \lambda (M_y\dot{M_z} \!-\! M_z\dot{M_y}), 
     \\
    \dot{M_y} \! = \!      ( \beta M_x                 \!-\! H_0   \!   -\! \gamma M_y)M_z                   \! +\!  \lambda (M_z\dot{M_x}\! -\! M_x\dot{M_z}), 
    \\
    \dot{M_z} \! = \!     H_0M_y \!   +\! \gamma (M_x^2\! +\! M_y^2)                 \!+\!    \lambda (M_x\dot{M_y} \!-\! M_y\dot{M_x})
	\end{cases} \!\!\!\!
    \label{DS}
\end{equation} 
 on the surface of the unit sphere.

 Once a fixed point ${\bf M}^{(0)}$ has been determined, we let ${\bf M}= {\bf M}^{(0)}+ \delta {\bf M}$,
 linearise the system \eqref{LL-components} 
 in $\delta {\bf M}$,  and consider solutions of the form
 \be
 \delta {\bf M}= {\bf m} \, e^{\mu t-ikx},
 \label{dm}
 \ee
 where ${\bf m}= (m_x, m_y, m_z)^T$ is a real constant vector and   $k$ a real wavenumber that may take values from $-\infty$ to $\infty$.
 We call the uniform static state unstable if at least one of the  roots $\mu$
 of
 the associated characteristic equation has 
  a positive real part in some interval of $k$.
Otherwise the state is deemed stable.  
 
 Since the equation \eqref{LLE} conserves the quantity ${\bf M}^2$, 
 the difference between $\left( {\bf M}^{(0)} \right)^2$ and the square of the vector  ${\bf M}^{(0)} + \delta {\bf M}$
 will be time-independent:
 \[
  \frac{\partial}{\partial t} \,
  \left( 2 \, \delta {\bf M} \cdot {\bf M}^{(0)}   \right) =0.
  \] 
  Substituting from \eqref{dm} and assuming $\mu \neq 0$,
 this gives
 \be
 {\bf m} \cdot {\bf M}^{(0)}=0.
 \label{mM}
 \ee
 Equation \eqref{mM} implies (a) that  the time-dependent perturbations of the uniform static states lie on the unit sphere;
 and (b) that 
 the characteristic equation 
  may not  have more than two nonzero roots, $\mu_1$ and $\mu_2$. 
  The third root ($\mu_3$)  has to be zero.   

 
 Apart from classifying stability  of the uniform static states, it is useful to know which of these solutions can serve as backgrounds to
 static magnetic solitons.
To weed out a priori unsuitable cases, we set
  $\mu=0$  in the characteristic equation and consider $k^2$ as a new unknown  (rather than a parameter that varies from $0$ to $\infty$).
 If all roots $(k^2)_n$ of the resulting equation are real positive,
 there can be no localised solutions asymptotic to the
  uniform static state  ${\bf M}^{(0)}$ as $x \to \pm \infty$. On the other hand, if there is at least one 
 negative or complex root, 
 the solution ${\bf M}^{(0)}$ remains a candidate for solitons'   background.

\subsection{Equatorial fixed points on the unit sphere}

\begin{figure}[t]
 \begin{center}
             \includegraphics*[width=0.99\linewidth]{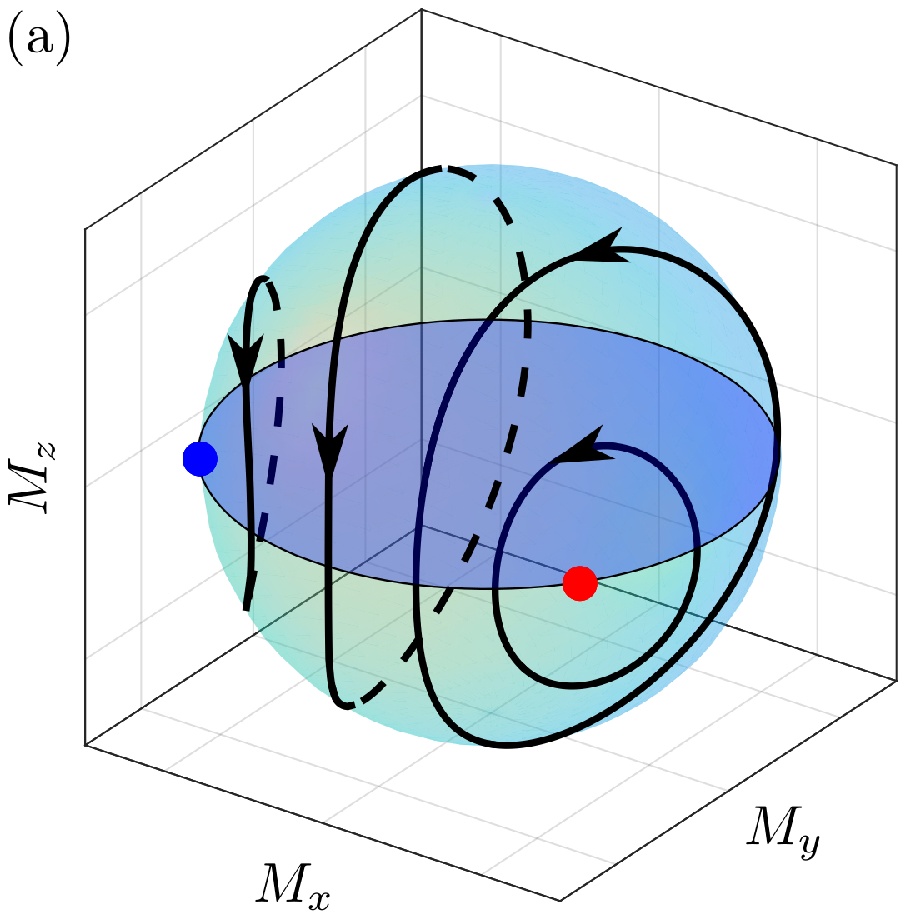}   
                \includegraphics*[width=0.99\linewidth]{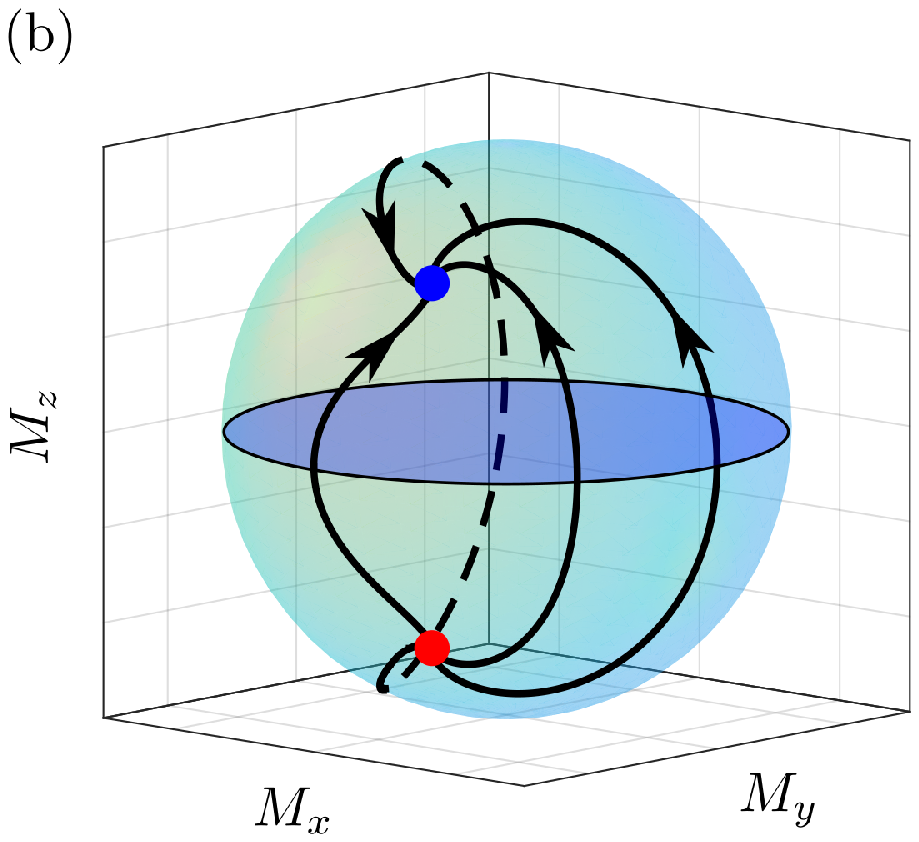}                           
                                   \caption{The phase portrait of   the dynamical system  \eqref{DS}  with
                                  $H_0>\sqrt{\beta^2+ \gamma^2}$ (a)  and $H_0< \gamma$ (b).
                                  In (a), two dots
                                  on the equator of the unit sphere  mark the fixed points of
                                  the vector field: the western (blue) and eastern point (red).  
                                  In (b), the blue dot indicates the northern and the red dot the southern
                                  fixed point. Apart from the fixed points, the figures show 
                                  a few representative trajectories; physically, these correspond to spatially-uniform 
                                  evolutions of magnetisation.
                                (The  portraits in (a) and (b)  are for $\lambda=0$.)
                       \label{pts}}
 \end{center}
 \end{figure}

One  family of  time-independent solutions of the system \eqref{DS}  describes 
a circle on the $(M_x,M_y)$-plane:
\[
M_x^2+  \left(M_y+ \frac{H_0}{2\gamma} \right)^2 =\frac{H_0^2}{4 \gamma^2},
\quad M_z=0.
\]
Imposing the constraint ${\bf M}^2=1$ leaves us with just  two members of the family:
\be
M_x^{(0)}=\pm \sqrt{1-\gamma^2/H_0^2},  \quad M_y^{(0)}=   -\gamma/H_0, \quad  M_z^{(0)}=0.
\label{B1}
\ee
In the system with $\gamma \neq 0$, these fixed points  are born
 as $H_0$ is  increased through  the value $H_0=\gamma$.
Since the points \eqref{B1} lie on the equator of the unit sphere, we will be referring 
to them simply as the {\it equatorial\/} fixed points, the eastern ($M_x^{(0)}>0$) and the western ($M_x^{(0)}<0$) one.
Fig \ref{pts}(a) depicts the equatorial fixed points in the phase portrait of the dynamical system \eqref{DS}. 


Linearising equation \eqref{LLE} about the uniform static state corresponding to an
 equatorial fixed point, 
we obtain two nonzero stability eigenvalues
\begin{subequations}        \label{EVtop}
 \begin{align}
        \mu_{1,2}
     =   \frac{ -\lambda (2K- \beta) \pm \sqrt{\lambda^2 \beta^2 - 4K(K-\beta)}}{2(1+\lambda^2)}, \label{EV1}   \\
     K= k^2+H_0 M_x^{(0)}.
    \label{EV2}
        \end{align}
        \end{subequations}
Making use of \eqref{EVtop} 
it is not difficult to see that in the easy-plane  or isotropic ferromagnet (i.e. in the  situation where $\beta \leq 0$),
 the eastern uniform static state ($M_x^{(0)}>0$)  is stable irrespective of the choice of $\gamma$, $\lambda$ and $H_0$.
On the other hand, when the anisotropy is easy-axis ($\beta>0$), the eastern state is stable if 
\be
H_0 \geq \sqrt{\beta^2+\gamma^2}
  \label{east_st}
\ee
 and unstable otherwise.

To check the suitability of the eastern uniform static state  as a background for solitons, we 
set $\mu=0$ in the expression \eqref{EV1}; this  transforms it into a quadratic  equation for $k^2$. When   $\beta < \sqrt{H_0^2-\gamma^2}$, 
both roots of this equation are negative:
$k^2=- \sqrt{H_0^2-\gamma^2}$ and $k^2= \beta-\sqrt{H_0^2-\gamma^2}$. This implies that there is a pair of exponentials $\exp ( -ik_{1,2}x )$ decaying to zero as $x \to -\infty$ and 
another pair decaying as $x \to +\infty$.
Therefore the uniform
static  state \eqref{B1} with $M_x^{(0)}>0$ can serve as a background to solitons
for any set of parameters  $\beta, \gamma, H_0, \lambda$ in its stability domain.

Turning to 
the west-point solution  ($M_x^{(0)}<0$  in  equation \eqref{B1}), a
simple analysis of  the eigenvalues \eqref{EVtop}  indicates that there are wavenumbers $k$ such that $\mathrm{Re} \, \mu>0$ for any quadruplets of 
 $\beta, \gamma,  H_0$  and $\lambda$.
Hence the western uniform static state  is always unstable. 
We are not considering it any further.

\subsection{Latitudinal fixed points} 
\label{LatStab}

  Another one-parameter family of constant solutions of the equation \eqref{LLE} 
forms a vertical straight line in the $M_x, M_y, M_z$-space:
\[
M_x= \frac{H_0 \beta}{\beta^2+\gamma^2}, \quad   M_y= -\frac{H_0 \gamma}{\beta^2+\gamma^2},    \quad   -\infty<M_z<\infty.
\]
The substitution of the above coordinates into ${\bf M}^2=1$ selects  two fixed points on the unit sphere:
\begin{align}
M_x^{(0)}= \frac{H_0 \beta}{\beta^2+\gamma^2}, \quad   M_y^{(0)}= -\frac{H_0 \gamma}{\beta^2+\gamma^2},    \nonumber \\ M_z^{(0)} = \pm \sqrt{1-\frac{H_0^2}{\beta^2+\gamma^2}}.
\label{B2}
\end{align}
Since these points lie above and below the equatorial $(M_x, M_y)$-plane, 
we will be calling them  the {\it latitudinal\/} fixed points:
 the northern ($M_z^{(0)}>0$) and the southern
($M_z^{(0)}<0$)  point. See Fig.\ref{pts}(b).

In the anisotropic equation ($\beta \neq 0$)
the northern and southern  points are born
as $H_0$ is decreased through $\sqrt{\beta^2+\gamma^2}$. 
In this case, there is a parameter interval 
$\gamma < H_0 < \sqrt{\beta^2+\gamma^2}$ where two pairs of fixed points, latitudinal and equatorial,  coexist.

The bifurcation diagram for the isotropic equation is different. When $\beta=0$, 
 the latitudinal fixed points 
 \eqref{B2} emerge as the  eastern and western points  \eqref{B1}
converge  and split out of the equatorial plane. 
In this case,   there is just one pair of uniform static states for any $H_0  \neq \gamma$: the equatorial pair for $H_0 > \gamma$ and the 
latitudinal pair for $H_0 <\gamma$.

 The
linearisation of equation \eqref{LLE}
about the  uniform static  state corresponding to the
 latitudinal fixed-point  \eqref{B2} gives
\be
\mu_{1,2}= 
-\frac{ \lambda Q  + \gamma M_z^{(0)}   \pm \sqrt{ \frac14 \lambda^2 \beta^2  h^2-P (P -\beta  h)}}
{1+\lambda^2},
  \label{Y1}
\ee
where
\[
P = k^2+\beta  -  \gamma \lambda M_z^{(0)},  \quad
Q= k^2+ \beta \left(1-\frac{h}{2}\right)     \label{Q}
\]
and
\be
h= \frac{H_0^2}{\beta^2+\gamma^2}.
\label{h}
\ee

A simple analysis demonstrates that when $\beta  \geq 0$, the north-point solution ($M_z^{(0)}>0$ in \eqref{B2}) 
is stable regardless of the values of $\lambda \geq 0$, $H_0$ and $\gamma>0$. 
As for the easy-plane anisotropy ($\beta<0$), the northern uniform static state is stable only when the inequalities $\lambda \leq \lambda_c$ and $H_0   \leq H_c$
are satisfied simultaneously.
Here
\be
\lambda_c=    \frac{\gamma}{|\beta|}  \frac{\sqrt{1-h }} { 1-  h/2}
\label{in1}  \ee
and
\be
H_c=
 \sqrt{  \frac{2 \gamma (\beta^2+\gamma^2)}{\beta^2}  \left( \sqrt{\beta^2+\gamma^2}-\gamma \right) }.
\label{in2} 
\ee
Note that $H_c$ is smaller than  $\sqrt{\gamma^2+\beta^2}$; hence the region $H_0 \leq H_c$ lies entirely within the northern point's  existence domain 
(defined by the inequality $H_0< \sqrt{\gamma^2+ \beta^2}$).

Finally, we consider the eigenvalues pertaining to 
 the southern uniform static state ($M_z^{(0)}<0$        in \eqref{B2}).
Our conclusion here is that in the isotropic and easy-plane ferromagnet ($\beta  \leq 0$),  this solution is unstable regardless of the choice of other parameters.
In the easy-axis situation ($\beta   >0$), 
the southern state is stable if  $\lambda  \geq \lambda_c$ with $\lambda_c$  as in \eqref{in1} ---
and unstable otherwise.

To determine the parameter region where 
the north-  and south-point  solutions   can serve as  backgrounds for solitons, we set $\mu=0$ in \eqref{Y1}. In each of the two cases, the resulting quadratic equation for $k^2$ 
has two positive roots only if $\beta<0$ is satisfied along with the inequality $H_0>H_c$, where $H_c$ is as in \eqref{in2}. This is the only no-go region for solitons.
Outside this region, the quadratic equation has either two negative or two complex roots; the corresponding uniform static states can serve 
as solitons' asymptotes. 

The bottom line is that either of the two latitudinal uniform static states  is suitable as a background  for solitons in its entire  stability  domain.

\subsection{Summary of uniform static states}

For convenience of the reader,  the stability properties of the constant solutions corresponding to the four fixed points are summed up
in Table \ref{stability-table}. 

Before turning to the perturbations of these uniform static states, it is worth noting their 
 symmetry properties.
Each of the equatorial states is \PT-symmetric in the sense that each of these two solutions is invariant under the product 
of the transformations \eqref{P} and \eqref{T}. 
In contrast, neither of the two latitudinal states is invariant; the {\PT \/} operator maps the northern solution to southern and the other way around.
The different symmetry properties of the equatorial and longitudinal solutions will give rise to different  invariances of equations for their small perturbations.

\begin{table}[h]
\setlength{\tabcolsep}{3pt}
\resizebox{\textwidth}{!}{
\begin{tabular}{|c||c|c|c|}
\hline
Fixed-point & \multirow{2}{*}[-2pt]{$\beta<0$}
	& \multirow{2}{*}[-2pt]{$\beta=0$} 
	& \multirow{2}{*}[-2pt]{$\beta>0$} \\[2pt] 
solution: & & & \\[2pt] \hline
\multirow{2}{*}[-2pt]{\bf eastern} 
	& \multirow{2}{*}[-2pt]{stable} 
	& \multirow{2}{*}[-2pt]{stable} 
	& stable if \\[2pt] 
	& & & $H_0 \ge \sqrt{\beta^2+\gamma^2}$ \\[2pt] \hline
\multirow{2}{*}[-2pt]{\bf western} 
	& \multirow{2}{*}[-2pt]{unstable} 
	& \multirow{2}{*}[-2pt]{unstable} 
	& \multirow{2}{*}[-2pt]{unstable} \\[2pt]
	& & & \\[2pt] \hline
\multirow{2}{*}[-2pt]{\bf northern} 
	& stable if $\lambda \le \lambda_c$ 
	& \multirow{2}{*}[-2pt]{stable} 
	& \multirow{2}{*}[-2pt]{stable} \\[2pt] 
	& and $H_0 \le H_c$ & & \\[2pt] \hline
\multirow{2}{*}[-2pt]{\bf southern} 
	& \multirow{2}{*}[-2pt]{unstable} 
	& \multirow{2}{*}[-2pt]{unstable}
	& \multirow{1}{*}{stable} \\[2pt]
	& & & if $\lambda \ge \lambda_c$ \\[2pt]
\hline
\end{tabular}}
\caption{Stability  of four  constant solutions of equation~\eqref{LLE}.}
\label{stability-table}
\end{table}

\section{Slow dynamics near bifurcation points}  
\label{NLS_dyn}

\subsection{Perturbation of equatorial fixed point} 
\label{Slow1}

Consider the eastern point of the pair of equatorial fixed points  \eqref{B1}:
\be
{\bf M}^{(0)}=  \left( \sqrt{1-\frac{\gamma^2}{H_0^2}},   -\frac{\gamma}{H_0}, 0 \right).
\label{east}
\ee
We assume that the parameters $\beta$, $\gamma$, $\lambda$ and $H_0$ lie in the stability domain of the uniform static state \eqref{east}. 

The plane orthogonal to the vector ${\bf M}^{(0)}$ is spanned by the vectors
 \[
{\bf A}= (0, 0, 1), \quad   {\bf B} =  \left( \frac{\gamma}{H_0},   \sqrt{1-\frac{\gamma^2}{H_0^2}},   0 \right).
\]
The unit vector $\bf M$ can be expanded over the orthonormal triplet 
 $\{\bf A, \bf B,  \bf M^{(0)} \}$:
 \[
 {\bf M}= \eta {\bf A} + \xi {\bf B} + \chi {\bf M^{(0)}}.
 \]
Letting  ${\bf M}(x,t) \to {\bf M}^{(0)}$ as $x  \to \pm \infty$, the coefficient fields $\eta$, $\xi$ and $\chi$ have the following asymptotic behaviour:
 \[
 \eta \to 0,   \  \xi \to 0, \ \chi \to 1 \ \mbox{as}  \ |x| \to \infty.
 \]
 
 The complex field $\Psi=\xi+i \eta$ satisfies
 \begin{align}
 i \dot \Psi= 
  \chi \Psi^{\prime \prime}-\Psi \chi^{\prime \prime} 
 +\lambda ( \Psi \dot \chi   -  \chi \dot \Psi  ) 
 - \sqrt{H_0^2-  \gamma^2}\Psi     \nonumber \\
  + \gamma (  \chi - i \eta \xi -1+\eta^2)
  + i \beta \eta \chi, 
 \label{long}
 \end{align}
 where $\chi= \sqrt{1-|\Psi|^2}$ while the prime and overdot indicate the derivative with respect to $x$ and $t$, respectively. 
Note that when $\lambda=0$, the equation  \eqref{long} is $\mathcal{PT}$-symmetric, that is, invariant under a composite transformation 
consisting of three involutions:
 $t \to -t$,  $x \to -x$, and $\Psi \to \Psi^*$.

Assume that $H_0$ is close to the bifurcation point of the uniform static state \eqref{east} --- that is, $H_0$ is slightly greater than $\gamma$.
In this case, $\Psi$ will depend on a hierarchy of slow times $T_n=\epsilon^n t$ and stretched spatial coordinates $X_n=\epsilon^{n/2} x$,
where $n=1,3,5,...$
and the small parameter $\epsilon$ is defined by
\[
 \epsilon^2=
1- \frac{\gamma^2}{H_0^2}.
\]
In the limit $\epsilon \to 0$ the new coordinates become independent so we can write
\begin{align*}
\frac{\partial}{\partial t}= \epsilon D_1 + \epsilon^3 D_3 + ...   \  ;  \\
\frac{\partial^2}{\partial x^2}= \epsilon \partial_1^2+ 2 \epsilon^2 \partial_1\partial_3 +\epsilon^3( \partial_3^2+ 2 \partial_1 \partial_5)+ ... \ ,
\end{align*}
where $D_n= \partial/ \partial T_n$ and $\partial_n= \partial/ \partial X_n$.
Assume, in addition,  that the anisotropy constant $\beta$ is of order $\epsilon$ and let
 $\beta= \epsilon \mathcal B$  with $\mathcal  B= O(1)$.
Considering small $\eta$ and $\xi$, we expand
\[
\Psi= \epsilon \psi_1 +  \epsilon^3  \psi_3 + ...  \ .
\]
Substituting the above expansions in \eqref{long}, we equate coefficients of like powers of $\epsilon$.  The order $\epsilon^2$ gives
a Ginsburg-Landau type of equation with a quadratic nonlinearity:
\be
(i +\lambda) D_1 \psi - \partial_1^2 \psi          + \frac{\gamma}{2} \psi^2             =   - \gamma \psi 
 +  \frac{\mathcal B}{2} (\psi-\psi^*).
\label{NLS_E}
\ee
(Here $\psi$ is just a short-hand notation for $\psi_1$.)

Note that in the derivation of \eqref{NLS_E} we took $\lambda$ to be $O(1)$.
If we, instead,  let $\lambda=O(\epsilon)$, the dissipative term would   fall out of the equation \eqref{NLS_E} and we would end up with a  nonlinear Schr\"odinger equation:
\be
i D_1 \psi - \partial_1^2 \psi   + \frac{\gamma}{2} \psi^2  =   - \gamma  \psi      +           \frac{\mathcal B}{2} (\psi-\psi^*).
 \label{NLS0}
\ee
The quadratic Schr\"odinger equation   \eqref{NLS0} does not have the U(1) phase invariance. However, 
 the equation is $\mathcal{PT}$-symmetric, that is, invariant under the  composite map
 $t \to -t$, $x \to -x$,  $\psi \to \psi^*$.
As we will see in section \ref{qua}, this discrete symmetry  is enough to stabilise solitons.

 \subsection{Perturbation of latitudinal fixed points}

 Choosing the background in the form of one of  the two latitudinal fixed points 
 \be
 {\bf M}^{(0)} = \left( \frac{\beta}{H_0} h, -\frac{\gamma}{H_0}h,  \, \pm \sqrt{1-h} \right),
 \label{latte}
 \ee
 we let ${\bf M}(x,t)$ approach the same point ${\bf M}^{(0)}$ as $x \to \pm \infty$. 
 In \eqref{latte}, $h$  is defined by  the equation \eqref{h}.

 As in the previous subsection, we 
 expand the magnetisation vector  over an orthonormal basis $\{ {\bf A}, {\bf B}, {\bf M}^{(0)} \}$:
\be
{\bf M}= \eta {\bf A}  +  \xi {\bf B}  + \chi {\bf M}^{(0)},
\label{MAB}
\ee
where, this time, 
 \[
 {\bf A}= 
  \left(
 \mp \frac{\beta}{H_0}\sqrt{h(1-h)},  \pm \frac{\gamma}{H_0} \sqrt{h(1-h)},  \sqrt{h} \right)
 \]
 and
 \[
 {\bf B}=
 \left(
 \frac{\gamma}{H_0} \sqrt{h},  \frac{\beta}{H_0} \sqrt{h}, 0  \right).
 \]

 We assume that $H_0$ is close to the  bifurcation point where 
 the northern and southern fixed points are born  (that is, $H_0$ is slightly smaller than $\sqrt{\beta^2+\gamma^2}$)
  and define a small parameter $\epsilon$:
 \[
 h= 1 -\epsilon^2.
 \]
As in the analysis of the equatorial fixed points,   we let  $\beta = \epsilon \mathcal B$,
 where $\mathcal B=O(1)$. 
 Assuming that  the magnetisation $\bf M$ is just a small perturbation of ${\bf M}^{(0)}$, we expand the small coefficients in \eqref{MAB}  in powers of $\epsilon$:
 \[
 \eta= \epsilon \eta_1+ \epsilon^3 \eta_3+ ...,
 \quad 
 \xi= \epsilon \xi_1+ \epsilon^3 \xi_3 + ... \ .
 \]
 The constraint $\eta^2+ \xi^2+\chi^2=1$ implies then 
 \[
 \chi= 1- \epsilon^2 \frac{\eta_1^2+\xi_1^2}{2} + ...  \ .
 \]

Substituting these expansions in the Landau-Lifshitz equation \eqref{LL-components}
and equating coefficients of like powers of $\epsilon$,  the order $\epsilon^2$ gives
\[
 D_1 \xi_1=  \partial_1^2 \eta_1 - \lambda   D_1 \eta_1-
\gamma (\eta_1 \xi_1 \pm \xi_1)
\]
 and
  \[
 D_1 \eta_1=  \lambda  D_1 \xi_1-  
  \partial_1^2 \xi_1 + \mathcal B \xi_1 \mp \gamma \eta_1 -
 \frac{\gamma}{2} (\eta_1^2-\xi_1^2).
 \]
  The above two equations can be combined into a single equation for the complex function
  $\psi=\xi_1+i \eta_1$:
\be
(i+ \lambda) D_1  \psi - \partial_1^2  \psi +\frac{\gamma}{2} \psi^2= \mp i \gamma \psi -\frac{\mathcal B}{2} (\psi+\psi^*). 
\label{NLS_NS}
\ee

The Ginsburg-Landau equation \eqref{NLS_NS} resembles the equation \eqref{NLS_E} governing the dynamics 
near the equatorial uniform static state;  however there is an important difference.
Namely, even if we let $\lambda=0$ in \eqref{NLS_NS}
[that is,   even if we assume that the damping is $O(\epsilon)$ or weaker  in the Landau-Lifshitz-Gilbert equation \eqref{LLE}], 
the resulting nonlinear Schr\"odinger equation will {\it not\/} become $\mathcal{PT}$-symmetric.
This fact will have important repercussions for the stability of solitons.

\section{Soliton excitations of equatorial state} 
\label{qua}

Letting 
\be
u(x,t)       =-   \frac13  \psi(X_1,T_1), \quad    
x= \frac{\sqrt \gamma}{2} X_1,
\quad
t= \frac{\gamma}{4} T_1,
\label{tranny}
\ee
 the Ginsburg-Landau equation \eqref{NLS_E} is cast in the form
\be
(i+\lambda)  u_t  -u_{xx} -6 u^2         =   - 4u  + b (u-u^*),
\label{fin}
\ee
where $b = 2 \mathcal B/ \gamma$.
(We alert the reader  that the scaled variables $x$ and $t$ do not coincide with the 
original $x$ and $t$ of the Landau-Lifshitz equation \eqref{LLE}. 
We are just re-employing the old symbols in a new context here.)

In the present section we consider localised solutions of the equation \eqref{fin} approaching 0 as $|x| \to \infty$.
Regardless of $\lambda$, the zero solution is stable if $b \leq 2$ and unstable otherwise. 
This inequality agrees with the stability range \eqref{east_st} of the eastern  uniform  static state within the 
original Landau-Lifshitz equation.
(Note that the term $b u^*$  plays the role of the 
parametric driver  in \eqref{fin} \cite{parametric};  the above stability criterion
states that the zero solution cannot sustain drivers with amplitudes greater than $b=2$.)

 \subsection{Fundamental soliton and its stability}

  \begin{widetext}
 
   \begin{figure}
 \begin{center}
                          \includegraphics*[width=0.49\linewidth]{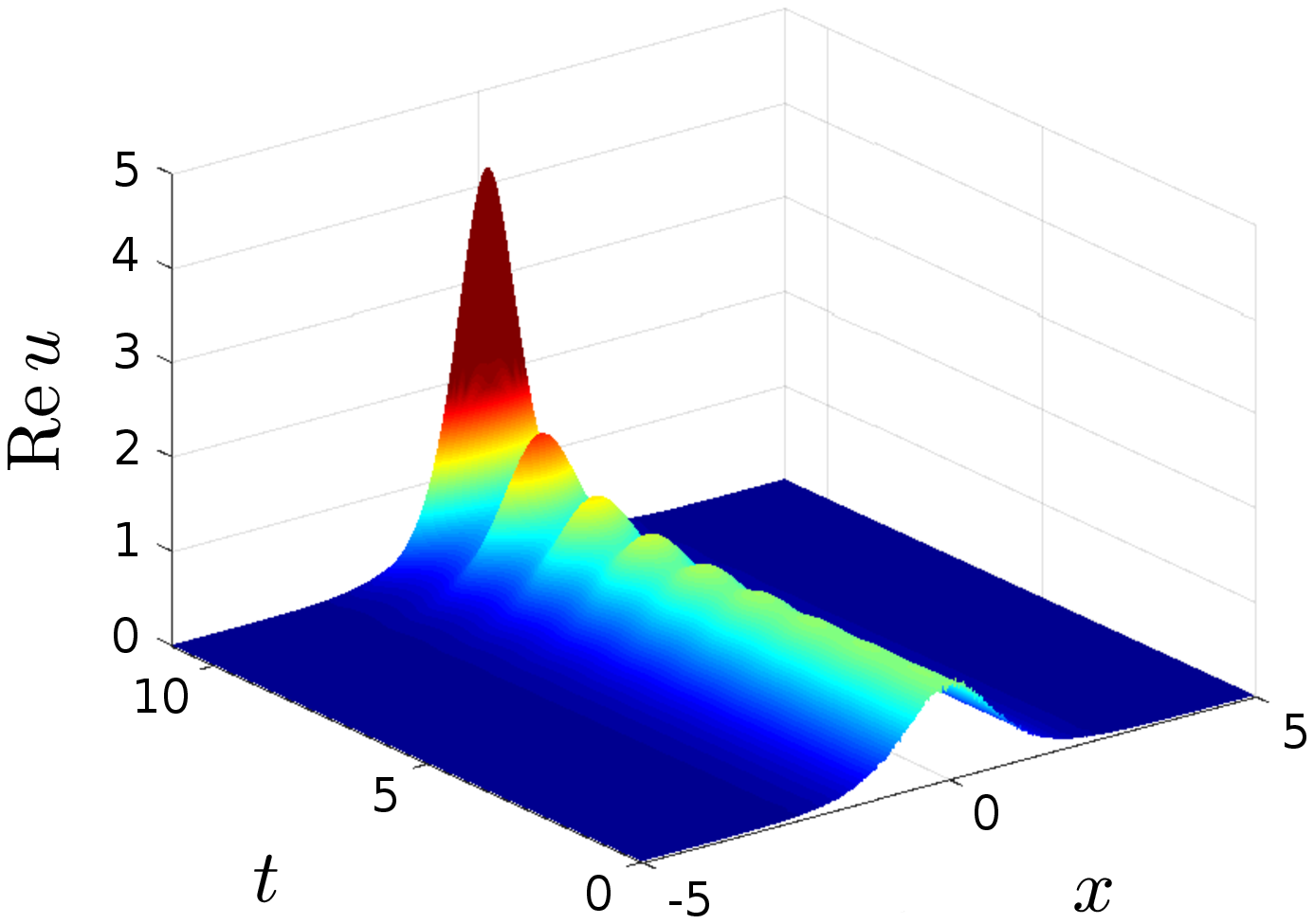}
              \includegraphics*[width=0.49\linewidth]{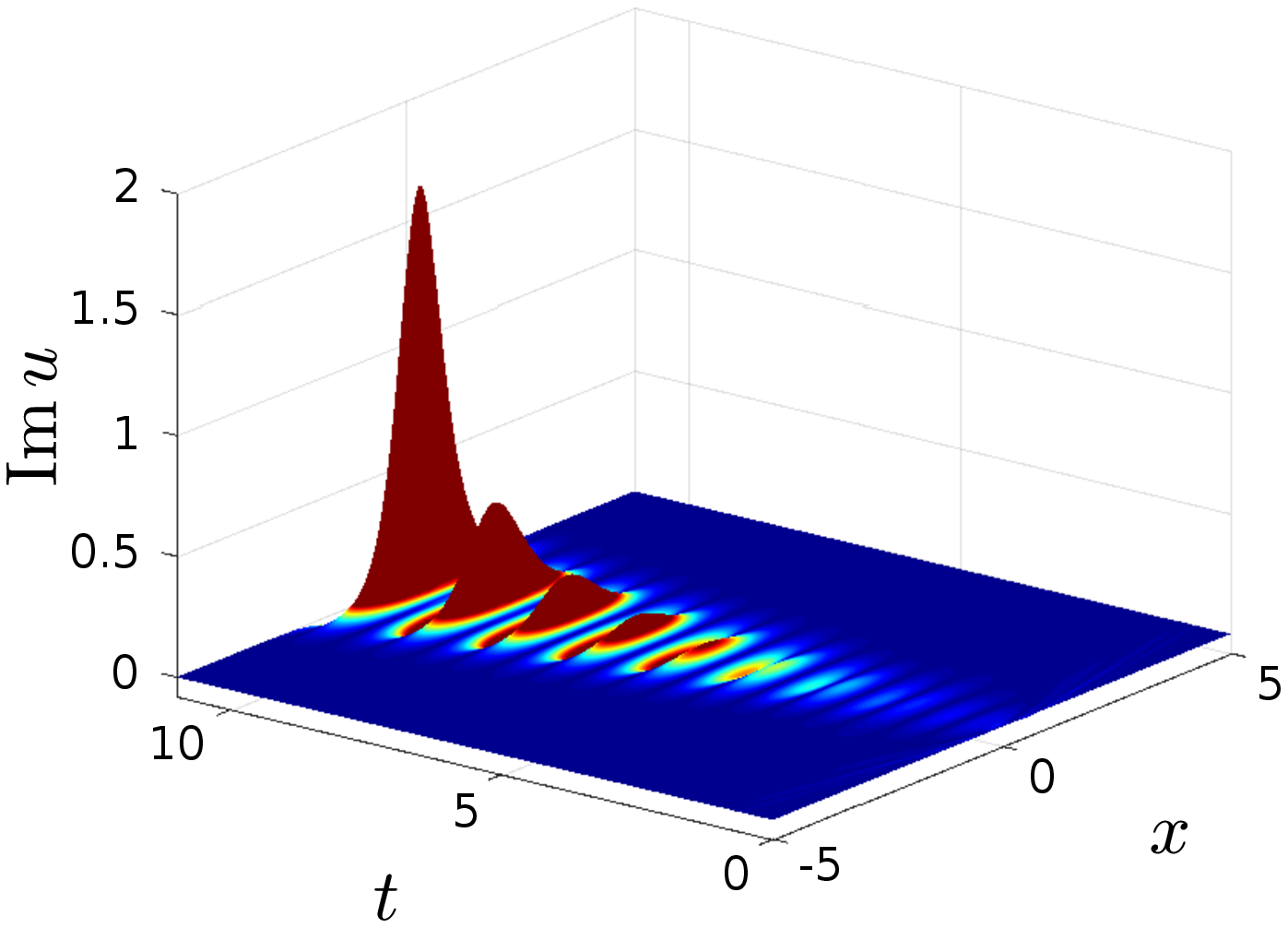}
                                   \caption{Instability of the fundamental soliton in the presence of damping. This evolution was obtained by the 
                                          direct numerical simulation of the  equation \eqref{fin}
                     with $b=0$ and  $\lambda=0.1$. The initial condition was in the form of the soliton \eqref{soliton} perturbed by 
                     a  random perturbation within  5\% of the soliton's amplitude. The  spatial interval of simulation was $(-58,58)$; in the plot it has been cut down  for visual clarity.
                       \label{inst_sech}}
 \end{center}
 \end{figure}  
 \end{widetext}

Equation \eqref{fin} has a stationary soliton solution:
\be
u_s=   \mathrm{sech}^2 x.
\label{soliton}
\ee 
To distinguish it from  localised modes with internal structure, 
  we refer to this solution as the  \textit{fundamental  soliton} --- or simply {\it sech mode}.
 Letting 
\[
u(x,t)=u_s(x)+ \varepsilon [f(x)+ig(x)] e^{\mu t}
\]
and linearising in small $\varepsilon$, we obtain an eigenvalue problem
\begin{subequations}    \label{vec}
\begin{align}
\mu ( g- \lambda f) = \mathcal H f, \\
-          \mu   ( f - \lambda g)= (\mathcal H - 2b) g,
\end{align}
\end{subequations}
with the operator 
\be
\mathcal H= - d^2/dx^2+4 - 12 \, \mathrm{sech}^2 x.
 \label{PT}
\ee

 The vector eigenvalue problem \eqref{vec} is reducible to a scalar eigenvalue problem of the form
\[
(\mathcal H - b+ \mu \lambda)^2  g    +( \mu^2-b^2) g =0.
\]
The stability exponents $\mu$ are roots of the quadratic equation
\[
 (E- b+ \mu \lambda)^2   +  \mu^2        - b^2          =0,    
 \]
 where $E$ is an eigenvalue of the operator $\mathcal H$: $\mathcal  H y=Ey$.
 The two roots are
 \be
 \mu^{(\pm)}=\frac{\lambda (b-E) \pm \sqrt{\lambda^2 b^2 +E(2b-E)}}{1+\lambda^2}.
 \label{mus}
 \ee

 The eigenvalues of  the P\"oschl-Teller operator  \eqref{PT} are $E_0=-5$,
$E_1=0$, and $E_2=3$,  with the eigenfunctions 
$y_0= \mathrm{sech}^3 x$, $y_1= \mathrm{sech^2} x \tanh x$
and $y_2= \mathrm{sech} \, x \left( 1- \frac54 \mathrm{sech}^2 x \right)$,
respectively. The continuous spectrum occupies the semiaxis $E_{\mathrm{cont}} \geq 4$,
with the edge eigenfunction given by $y_3= \tanh x \left( 1-\frac53 \tanh^2 x \right)$.
For each eigenvalue $E_n$, $n=0,1,2$,   equation \eqref{mus} yields two roots, $\mu^{(+)}_n$ and $\mu^{(-)}_n$.

 In the analysis of the roots \eqref{mus} we need to distinguish between two situations: 
 damped ($\lambda>0$) and  undamped one ($\lambda=0$). 
  Assume, first, that $\lambda>0$ and let, in addition, $b \geq 0$.
 It is not difficult to check that   the root  $\mu_n^{(+)}$  will
have a positive real part 
 provided the corresponding eigenvalue $E_n$ satisfies
 $E_n < 2b$. On the other hand, 
 the set of three eigenvalues of the operator \eqref{PT} does include
a negative eigenvalue ($E_0$) that satisfies $E_0<2b$ regardless of the particular value of $b \geq 0$.
 Therefore
  the soliton has an exponent  $\mu_0^{(+)}$ with
$\mathrm{Re} \, \mu_0^{(+)} >0$ for any $b \geq 0$.

 In the case where $\lambda>0$ but $b < 0$,  the root  $\mu_n^{(+)}$  will
have a positive real part 
 provided $E_n$ satisfies $E_n< 0$. As in the previous case, this inequality is satisfied by the eigenvalue $E_0$
 so that  the soliton has an exponent with $\mathrm{Re} \, \mu_0^{(+)} >0$ for any $b < 0$.

  We conclude that
 the fundamental soliton of the equation \eqref{fin}
is  unstable in the presence of damping --- regardless of the sign and magnitude of the anisotropy coefficient $b$.
 Figure \ref{inst_sech} illustrates the evolution of  a weakly perturbed soliton in the Ginsburg-Landau equation 
  with $\lambda \neq 0$.

 Turning to the situation with  $\lambda=0$ we assume, first, that 
 $b > 0$.  The equations \eqref{mus}
 will  give a pair of opposite real roots $\mu_n^{(\pm)}$ if the corresponding eigenvalue satisfies  $0<E_n < 2b$
 and a pair of pure imaginary roots otherwise. 
  The only positive eigenvalue of the  operator \eqref{PT} is $E_2=3$; it satisfies the above inequality if $b> 3/2$. 
 
  In the situation where $\lambda=0$ but $b<0$, the pair of opposite exponents $\mu_n^{(\pm)}$ is real if $E_n$ falls in the interval 
 $2b<E_n<0$ and pure imaginary if $E_n$ lies outside  this interval. The only negative eigenvalue is $E_0=-5$; 
 it falls in the interval in question if $b< -5/2$. 
 
 Finally, in the isotropic ferromagnet ($b=0$) the stability exponents are all pure imaginary: $\mu_n^{(\pm)}= \pm i E_n$.

Combining the intervals where all exponents are pure imaginary  gives us the stability region of the undamped fundamental soliton
in terms of the anisotropy to  spin-current ratio:
\be
\label{stab_reg}
-\frac52  \leq b \leq \frac32.
\ee

 \subsection{Twisted modes in isotropic ferromagnet}

The Ginsburg-Landau equation \eqref{fin}  with $b=0$
admits an additional pair of  localised solutions:
\be
u_{\mathrm{T}}=2 \mathrm{sech}^2 (2x)  \pm 2i  \,\mathrm{sech} (2x) \tanh (2x).
\label{twist}
\ee
The modulus of $u_\mathrm{T}(x)$ is bell-shaped
while its phase grows or decreases by $\pi$  as $x$ changes from $-\infty$ to $+\infty$.
The solution  looks like a pulse twisted by $180^\circ$ in the $(\mathrm{Re} \,  u, \mathrm{Im}  \, u)$-plane. 
In what follows, we refer to each of equations \eqref{twist} as a {\it twisted}, or simply {\it sech-tanh}, mode. 

 \begin{widetext}
 
   \begin{figure}
 \begin{center}
                            \includegraphics*[width=0.49\linewidth]{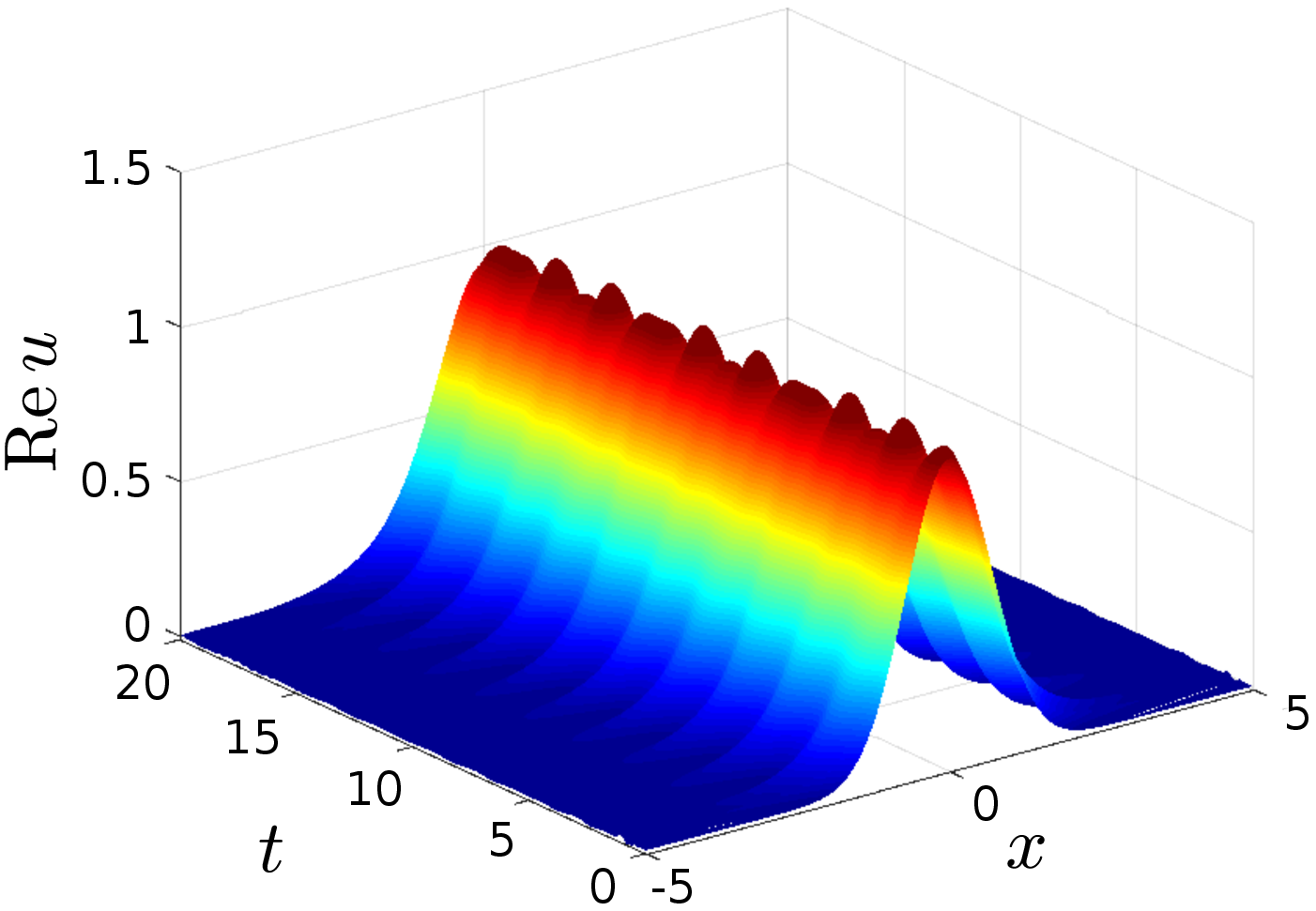}
                                                         \includegraphics*[width=0.49\linewidth]{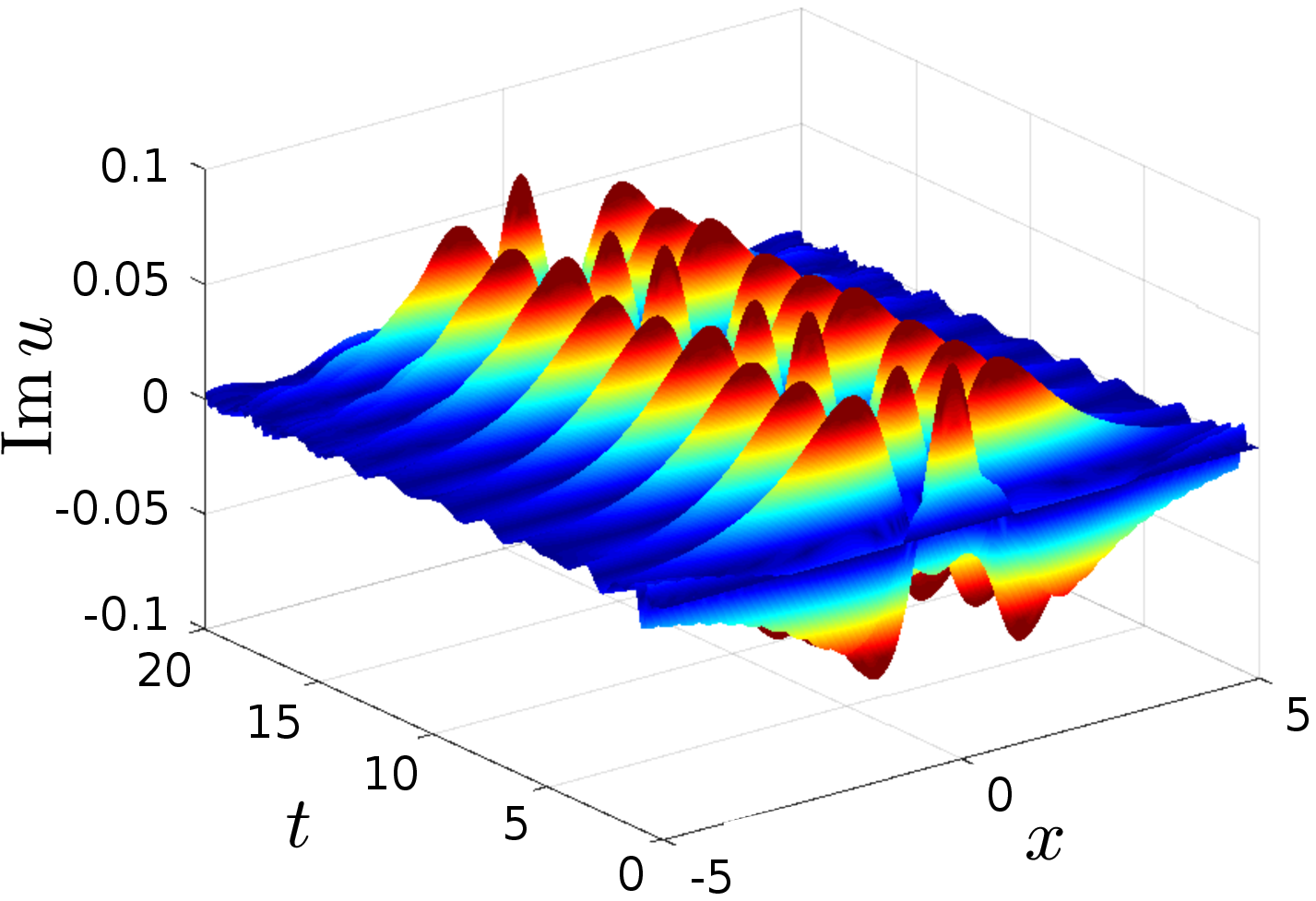}
                                                                                            \caption{The evolution of the initial condition in the form of a gaussian, $u(x,0)=\exp (-x^2)$,
                                   in the equation \eqref{fin} with $\lambda=0$ and $b=0$. Left panel: $\mathrm{Re} \, u$; right panel: $\mathrm{Im} \, u$.
                                   The emerging solution is a breather with a small imaginary part and the real part close to the 
                                    soliton \eqref{soliton}. Note that the figure shows only a portion  of the full simulation interval $(-58,58)$. 
                       \label{breather}}
 \end{center}
 \end{figure}  
 \end{widetext}

Linearising equation \eqref{fin} about the twisted mode \eqref{twist}
and assuming that the small perturbation depends on time as $e^{\mu t}$, 
we arrive at an eigenvalue problem
\be
\mathcal{L} f (X)= 
-  \frac{\mu}{4} (\lambda+i)  f (X)
 \label{EVt}
 \ee
 for the Schr\"odinger operator with the Scarff-II complex potential:
\be
\mathcal{L}= -\frac{d^2}{dX^2} +1  -  6 \mathrm{sech}^2 X  \mp  6 i  \,  \mathrm{sech} \, X \tanh X.
\label{Scarff} 
\ee
In \eqref{EVt}-\eqref{Scarff}, $X=2x$.

The  \PT-symmetric operator \eqref{Scarff} has an all-real spectrum including three discrete eigenvalues \cite{Ahmed}. 
Let $y_n$  be the eigenfunction associated with an eigenvalue $E_n$: $\mathcal L y_n=E_ny_n$.
The eigenvalue-eigenfunction pairs are then given by
\begin{gather} 
E_0=-\frac54, \quad y_0= (\mathrm{sech}^2 X \pm  i  \,\mathrm{sech} X \tanh X)^{3/2};
\nonumber \\
E_1=0, \quad y_1= \mathrm{sech} X (\mathrm{sech} X \pm  i \tanh X)^2,
\label{ScEV}
\end{gather}
and  $E_2= 3/4$ with
\be
y_2=(3\pm  2 i \sinh X)(\mathrm{sech}^2 X \pm  i   \, \mathrm{sech} X \tanh X)^{3/2}.
\label{ScEV2}
\ee

Each of the eigenvalues $E_n$ gives rise to a stability exponent 
\[
\mu_n=  4 \frac{i- \lambda}{1+\lambda^2} E_n
\]
in  equation \eqref{EVt}.
When the dissipation coefficient $\lambda>0$, the  exponent pertaining to the negative eigenvalue $E_0$ has a positive real part.
Accordingly, the twisted modes \eqref{twist} are unstable in the presence of damping.
In contrast, when $\lambda=0$, all exponents $\mu_n$ ($n=0,1,2)$ are pure imaginary so 
the twisted modes are  stable.

\subsection{Oscillatory modes}

An interesting question is whether there are any other stable localised structures --- in particular, 
 in the situation where the equation \eqref{fin} has zero damping.
Figure \ref{breather} illustrates the evolution of a gaussian initial condition 
$u(x,0)= \exp (-x^2 )$ that can be seen as a nonlinear perturbation of the  soliton \eqref{soliton}. 
The gaussian evolves into an oscillatory localised structure (a kind of a breather) which remains close 
to the soliton \eqref{soliton} --- but does not approach it as $t \to \infty$. 
This observation suggests that equation \eqref{fin} with $\lambda=0$ has a family of stable time-periodic spatially localised solutions,
with the stationary soliton \eqref{soliton} being just a  particular member of the family.

It is fitting to note that the existence of breather families is common to nonlinear \PT-symmetric equations \cite{breathers}.
Breathers prevail among the products of decay of generic localised initial
conditions \cite{breathers,PTsolitons}.

\subsection{Stable solitons in two dimensions}

   \begin{figure}[t]
 \begin{center}
        \includegraphics*[width=0.99\linewidth]{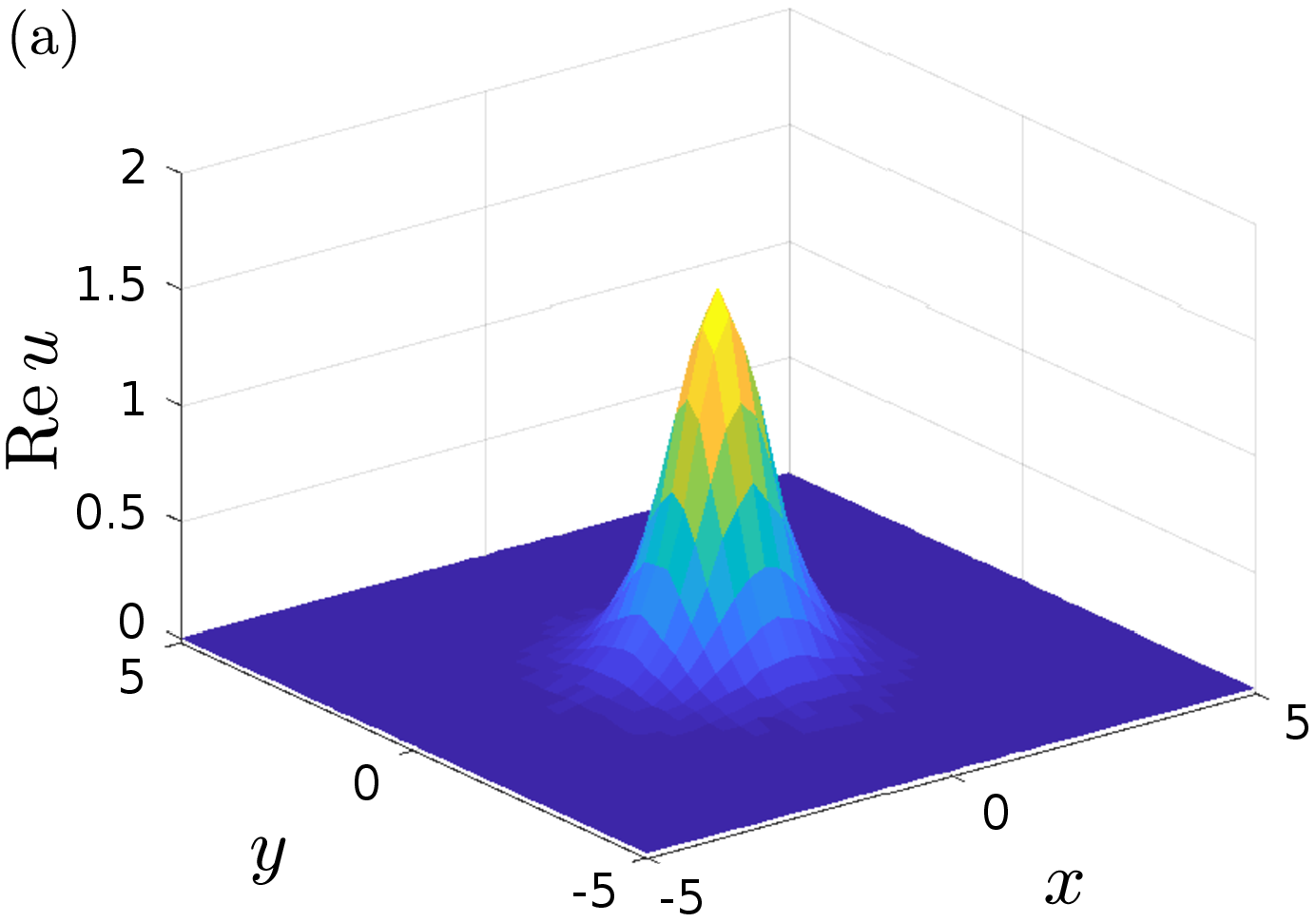}             \includegraphics*[width=0.99\linewidth]{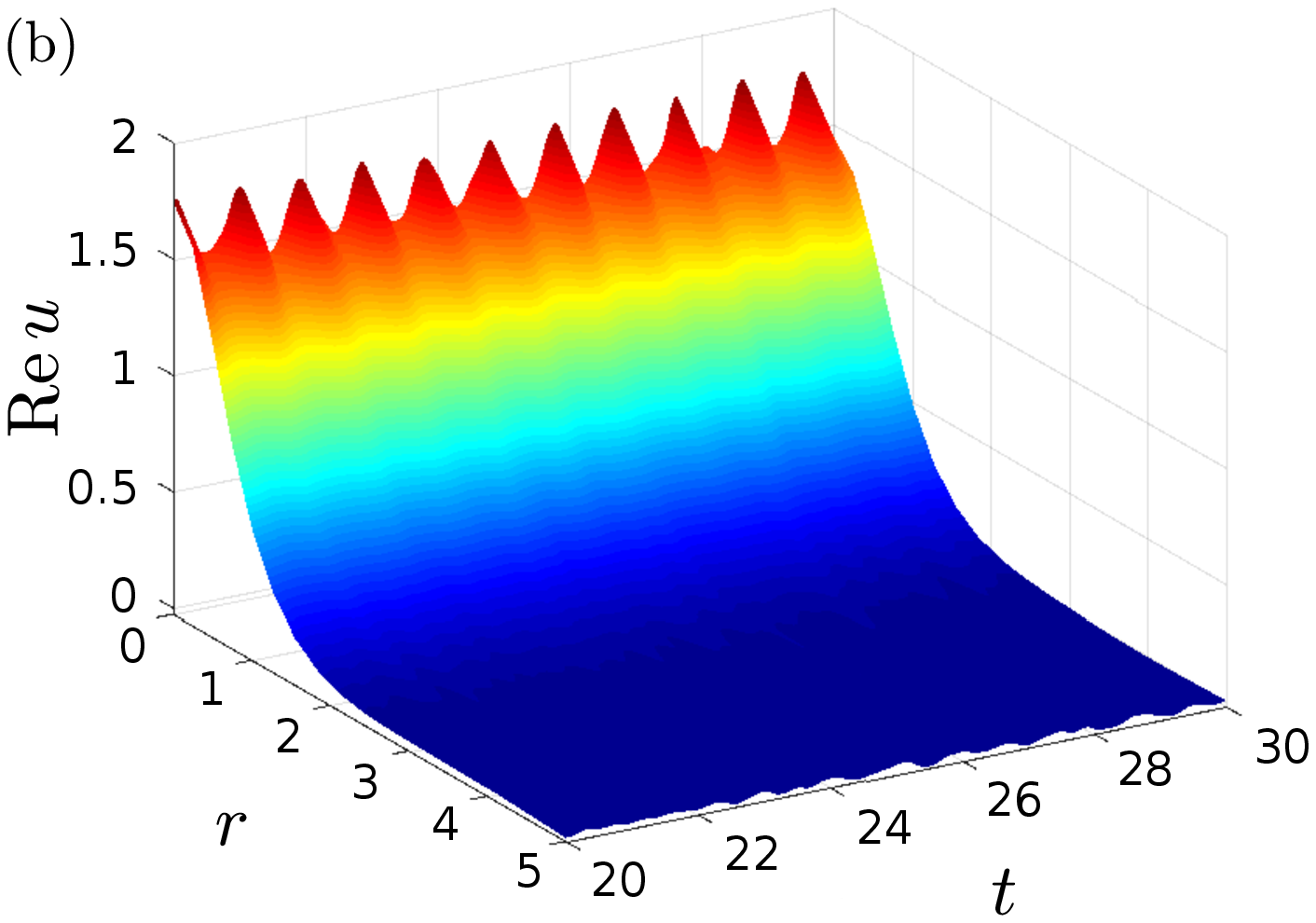}
                                  \caption{ Localised solutions of the  quadratic Schr\"odinger equation on the plane:
                                  the stationary soliton (a) and a breather (b). Both figures were produced by direct numerical simulations 
                                  of equation  \eqref{NLS2D} with $b=0$. 
                                  In panel (a),  the initial condition was taken in the form of the soliton \eqref{Rr}
                                  perturbed by 
                     a  random perturbation within  5\% of the soliton's amplitude. After $t=100$, the solution (shown in the panel)
                     remains close to the soliton.
                                                       In panel (b), the initial condition was chosen as $u=1.6 \, \exp (-r^2)$. After an initial transient, the 
                                                       solution settles to a localised oscillatory state shown in the figure.       
                       \label{2D}}
 \end{center}
 \end{figure}

We close this section with a remark on the Landau-Lifshitz-Gilbert-S\l{}onczewski equation in two dimensions:
\begin{align}
   \frac{\partial {\bf M}}{\partial t} =  - {\bf M}\times  \nabla^2  {\bf M}
     -  {\bf M}\times {\bf H}
    - \beta  (  {\bf M} \cdot    \hat  {\bf z} )  \,  {\bf M}    \times    \hat{\bf z}  \nonumber \\
       - \gamma {\bf M}\times {\bf M} \times \hat{\bf z} 
    + \lambda {\bf M}\times \frac{\partial {\bf M} }{\partial t}.
    \label{LLE2D}
\end{align}
Here 
$\nabla^2= \frac{\partial^2}{\partial x^2}  +  \frac{\partial^2}{\partial y^2}$. 
Assuming that $H_0$ is only slightly above $\gamma$ and that the anisotropy $\beta$ and damping $\lambda$ are small,
we consider a perturbation of the east-point uniform  state \eqref{east}.
Following the asymptotic procedure outlined in section \ref{Slow1}, the equation \eqref{LLE2D}
is reducible  in this limit to a  planar Schr\"odinger equation:
\be
 i u_t  = u_{xx} +u_{yy} + 6 u^2         - 4u  + b (u-u^*),
\label{NLS2D}
\ee
where
\[
b=  \frac{2H_0}{\gamma} \frac{\beta}{\sqrt{H_0^2-\gamma^2}}.
\]

Like its one-dimensional counterpart \eqref{NLS0}, equation \eqref{NLS2D} is \PT-symmetric. The \PT-operation can be chosen, for instance, 
in the form  
\[
t \to -t, \quad  x \to -x, \quad   y \to -y, \quad  u \to  u^*.
\]

The quadratic Schr\"odinger equation \eqref{NLS2D} has a static radially-symmetric soliton solution,
\be
u_s(x,y)= \mathcal R(r),   \label{Rr}
\ee
where $\mathcal R(r)$ is a nodeless (bell-shaped) solution of the boundary-value problem
\begin{gather*}
\mathcal R_{rr} + \frac{1}{r} \mathcal R_r           - 4\mathcal R            + 6 \mathcal R^2              =0, \\
\mathcal R_r(0)=0, \quad \mathcal R (r) \to 0 \ \mbox{as} \ r \to \infty.
\end{gather*}

Postponing the detailed stability analysis of the soliton \eqref{Rr} to future publications, 
we restrict ourselves to the simplest case of isotropic ferromagnet, $b=0$. 
A  numerical simulation of equation \eqref{NLS2D} with the initial condition in the form of 
the  noise-perturbed soliton \eqref{Rr} indicates that the soliton is stable against small perturbations.
[See Fig \ref{2D}(a).]  On the other hand,  generic localised initial conditions evolve into  time-periodic  breather-like states
[Fig \ref{2D}(b)]. This suggests that the quadratic Schr\"odinger equation \eqref{NLS2D} 
[and hence the planar Landau-Lifshitz equation \eqref{LLE2D}] supports a 
 broad class of stable stationary and oscillatory localised 
  structures.

\section{Soliton excitations of    latitudinal state}
\label{Latitudinal}

The scaling transformation \eqref{tranny}   takes
 the equations \eqref{NLS_NS} to the nondimensional form
\be
(i+\lambda) u_t  - u_{xx}   -  6 u^2 = \mp 4i u  - b (u +u^*).
\label{fin34}
\ee
 As in section \ref{qua}, $b = 2 \mathcal B/ \gamma$ here.

In what follows, we confine ourselves to the analysis of the isotropic equations
($b=0$) as it is the only regime where we were able to obtain soliton solutions of \eqref{fin34}. 
In the isotropic case, 
the $u=0$  solution of the top-sign equation   in \eqref{fin34}     is stable and that of the 
bottom-sign equation  unstable ---   regardless of whether $\lambda$ is zero or not.
(This agrees with the stability properties of the north and south  fixed-point  solutions of the Landau-Lifshitz equation; see section \ref{LatStab}.)
Hence we only keep the top-sign equation in what follows.


\subsection{ \textit{sech} mode}

Letting $b=0$, 
the top-sign equation  in  \eqref{fin34}
can be further transformed to
\be
 (1-i  \lambda) w_t= w_{zz}-4 w +6 w^2, 
\label{wt}
\ee
where 
\be
w(z,t)=-i     u, 
 \quad z= e^{ i \pi/4} x.
\label{trans} \ee
An obvious static solution of the equation \eqref{wt} is $w_s= \mathrm{sech}^2 z$; the corresponding solution of
the original  equation \eqref{fin34} is
\be
u_s(x)=  i \, \mathrm{sech}^2 \left( e^{  i \frac{\pi}{4}} x \right).
\label{sechcom}
\ee
The solution \eqref{sechcom} decays to zero as $x  \to  \pm \infty$ and does not have singularities on the real line.
Similar to the solution \eqref{soliton}   over the equatorial background,  we term the solution \eqref{sechcom} the {\it sech  soliton}.

To classify the stability of the soliton \eqref{sechcom}, we linearise equation   \eqref{wt} 
 about $w_s= \mathrm{sech}^2 z$.
Assuming that the small perturbation depends on time as $e^{\mu t}$, we obtain
\be
\mu = - \frac{1+i \lambda}{1+\lambda^2} E,
\label{mu}
\ee
where $E$ is an eigenvalue of the P\"oschl-Teller operator
\[
\mathcal H = -\frac{d^2}{dz^2}   +4 - 12  \,  \mathrm{sech}^2 z. 
\]
The operator acts upon functions $y(z)$
defined on the line $z=  e^{ i \pi/4} \xi$  ($-\infty< \xi < \infty$)  on the complex-$z$ plane
and satisfying the boundary conditions $y \to 0$ as $\xi \to \pm \infty$.

As discussed in the previous section, the equation $\mathcal H y=Ey$ with $E=-5$
 has a solution 
$y_0=  \mathrm{sech}^3 z$. The function $ \mathrm{sech}^3(  e^{i \pi/4} \xi  )$ is nonsingular for all $-\infty< \xi < \infty$ and decays to zero as
$\xi \to \pm \infty$; hence $E_0=-5$ is a discrete eigenvalue of the operator $\mathcal H$. 
The corresponding exponent $\mu$ in \eqref{mu} has a positive real part regardless of $\lambda$. 
This implies that the  {\it sech\/} soliton  \eqref{sechcom}   is unstable irrespective of whether $\lambda$ is zero or not.

\subsection{\textit{sech-tanh} modes}

Applying the transformation \eqref{trans} to 
 the  solutions
\be
w_\mathrm{T}  =2 \mathrm{sech}^2 (2z)   \pm  2i  \,\mathrm{sech} (2z) \tanh (2z)
\label{tw3} 
\ee
of
the equation  \eqref{wt}, we obtain
 a pair of localised solutions of the original equation \eqref{fin34}:
\be
u_\mathrm{T}= \mp 2 \,  \mathrm{sech} (2  e^{i\pi/4}x) \tanh (2 e^{i\pi/4}x)
+ 2i \,  \mathrm{sech}^2 (2 e^{i\pi/4}x).
\label{tw4}
\ee
By analogy with solutions \eqref{twist} over the equatorial background, we are referring to
\eqref{tw4} as the \textit{sech-tanh modes}.

Linearising equation \eqref{wt} about its  stationary solutions \eqref{tw3}
and assuming that the small perturbation depends on time as $e^{\mu t}$, we 
obtain the following equation for the exponent $\mu$:
\[
\mu= - 4 \frac{1+i \lambda}{1+\lambda^2} E.
\]
Here $E$ is an eigenvalue of the Scarff-II operator: 
\begin{subequations}\label{Sc}
\begin{gather}
\mathcal L y= E y, \\
\mathcal L=  -\frac{d^2 }{dZ^2} +1  -6 \, \mathrm{sech}^2 Z  \mp 6i   \, \mathrm{sech} Z \tanh Z,
\end{gather}
\end{subequations}
with $Z=2z$. 
The eigenvalue problem \eqref{Sc} is posed 
 on the line 
 \be Z= e^{i \pi/4}  \xi, \quad   -\infty< \xi< \infty     
 \label{line}
 \ee
  on the complex-$Z$ plane,
 with the boundary conditions 
$y \to 0$  as $\xi \to \pm \infty$. 

Three solutions of the equation \eqref{Sc} are in \eqref{ScEV}-\eqref{ScEV2}.
Since $y_n(Z)$ ($n=0,1,2$) are nonsingular and decay to zero as $Z$  tends to infinity in either direction along the line \eqref{line},
these solutions are eigenfunctions of the operator $\mathcal L$ --- and the corresponding $E_n$ are eigenvalues.
The exponent $\mu_0$ pertaining to the eigenvalue $E_0=-5/4$ has a 
 positive real part:
\[
\mu_0=  -4 \frac{1+i \lambda}{1+\lambda^2} E_0.
\]
Consequently, the \textit{sech-tanh}  modes \eqref{tw4} are unstable --- no matter whether $\lambda$ is zero or not.

\subsection{Summary of one-dimensional solitons}

The stability properties of  six localised  modes supported by the quadratic Ginsburg-Landau equations \eqref{fin} and \eqref{fin34}
are 
summarised in Table    \ref{soliton-stability-table}.
The Table includes two 
\textit{sech} solitons (the fundamental soliton \eqref{soliton} and its latitudinal-background counterpart, equation \eqref{sechcom})
and four \textit{sech-tanh} modes
(the twisted modes \eqref{twist} and their latitudinal  analogs \eqref{tw4}).

\begin{table}[h]
\setlength{\tabcolsep}{3pt}
\resizebox{0.85\textwidth}{!}{
\begin{tabular}{|c||c|c|}
\hline
\multirow{3}{*}[-4pt]{\begin{tabular}{@{}c@{}}Nonlinear \\ mode\end{tabular}} & 
	\multirow{3}{*}[-4pt]{\begin{tabular}{@{}c@{}}over equatorial \\ background\end{tabular}} & over latitudinal \\[2pt]
& & background \\[2pt]
& & (with $b=0$) \\[2pt]
\hline
\textit{\textbf{sech}} & stable if $\lambda=0$ & \multirow{2}{*}[-2pt]{unstable} \\[2pt]
{\bf soliton} & and $-\frac{5}{2} < b < \frac{3}{2}$ & \\[2pt] \hline
\textit{\textbf{sech-tanh}} & exist if $b=0$; & \multirow{2}{*}[-2pt]{unstable} \\[2pt]
{\bf modes} & stable if $\lambda=0$ & \\[2pt]\hline
\end{tabular}}
\caption{Stability of  the stationary  nonlinear modes in one dimension. The middle column classifies solutions of the equation \eqref{fin}
while the right-hand column corresponds to solutions of  \eqref{fin34}.
}
\label{soliton-stability-table}
\end{table}

\section{Concluding remarks} 
\label{Conclusions}


We have studied nonlinear structures associated with 
the spin torque oscillator --- an 
open system described by the Landau-Lifshitz-Gilbert-S\l{}onczewski equation. 
In the limit of zero damping 
($\lambda=0$),  this nonconservative system is found  to be \PT-symmetric.
The {\it nearly\/}-\PT  \/ symmetric equation
 corresponds to small nonzero $\lambda$.
In this paper, we have considered both nearly-symmetric and nonsymmetric 
oscillators (small and moderate $\lambda$).

The spin torque oscillator has four stationary states of  uniform magnetisation;
they are described by four fixed points on the unit ${\bf M}$-sphere.
 Two of these states have their magnetisation vectors lying in the equatorial plane of the unit sphere while the other two correspond to fixed points in the northern and southern hemisphere,
 respectively.  
 We have assumed  that the external magnetic field $H_0$ has been tuned to  values  $\epsilon^2$-close to   the bifurcation
 points of the ``equatorial" and ``latitudinal" uniform static states, 
 and that 
 the ferromagnet is only weakly anisotropic: $\beta=O(\epsilon)$.
In that limit, small-amplitude localised perturbations of the uniform static states satisfy the Ginsburg-Landau equations  --- equations \eqref{fin} and \eqref{fin34}, respectively.

If the damping coefficient
 $\lambda$ is $O(\epsilon)$ or smaller,
   each of 
   the two Ginsburg-Landau reductions becomes a quadratic nonlinear Schr\"odinger equation. 
   Of the two Schr\"odinger  equations,    
  the  one corresponding to perturbations of the ``equatorial" uniform static state
    turns out  to be \PT-symmetric. (Thus the asymptotic reduction of a
    {\it nearly\/}  \PT-symmetric Landau-Lifshitz system  is  {\it exactly\/} \PT-symmetric.)
 This Schr\"odinger equation 
    proves to be quite remarkable.
 Indeed, despite both our Ginsburg-Landau    reductions  supporting soliton solutions, it is 
 only in the \PT-symmetric Schr\"odinger  limit that the solitons are found to be stable.

   The \PT-symmetric Schr\"odinger equation supports two types of stable solitons. 
The constant-phase solution \eqref{soliton} is stable in a band of $\beta=O(\epsilon)$ values,
extending from the easy-axis to the easy-plane region.  [The stability band is demarcated by the inequality \eqref{stab_reg}.]
On the other hand, a pair of stable solitons with the twisted phase, equations \eqref{twist}, are only supported
by the nearly-isotropic ferromagnet: $\beta=O(\epsilon^2)$ or smaller. 
In addition to stable static solitons,  the \PT-symmetric Schr\"odinger equation exhibits  stable breathers.
 
 In the two-dimensional geometry, the Landau-Lifshitz equation for the spin torque oscillator 
 admits an asymptotic reduction  
 to a planar quadratic 
 Schr\"odinger equation,  equation \eqref{NLS2D}. Like its one-dimensional
 counterpart, the   \PT-symmetric planar Schr\"odinger equation 
 has stable static and oscillatory soliton solutions. 
 
 Finally, it is worth re-emphasising here that the \PT-symmetric Schr\"odinger equation is a 
 reduction of the whole family of 
  nearly-{\PT\/} symmetric Landau-Lifshitz-Gilbert-S\l{}onczewski equations with  $\lambda =O(\epsilon)$ ---
  and not just  of its  special case with $\lambda=0$.
  Therefore our conclusion on the existence of stable solitons 
  is applicable  to the physically relevant class of spin torque oscillators with nonzero damping. \\

\acknowledgments


We thank  Boris Ivanov  and  Andrei Slavin 
for useful discussions.
This project was supported by the NRF of South Africa (grants No  105835, 120844 and 120467).

\

\end{document}